\documentclass{article}
\usepackage[utf8]{inputenc}
\usepackage{todonotes}
\usepackage{amsthm,amsmath,euscript,mathrsfs,multirow,amsfonts}

\usepackage{graphicx}
\usepackage{stackengine}
\usepackage[top=1in,bottom=1in,left=1in,right=1in]{geometry}
\usepackage[square,numbers,sort&compress]{natbib}
\usepackage{subcaption}

\def\smcpp{SMC\texttt{++} }

\usepackage{authblk}
\title{Inference of Population History using Coalescent HMMs:\\ Review and Outlook}

\author[a]{Jeffrey P. Spence}
\author[b]{Matthias Steinr\"ucken}
\author[c]{Jonathan Terhorst}
\author[d,e,*]{Yun S. Song}

\affil[a]{Computational Biology Graduate Group, University of California, Berkeley}
\affil[b]{Department of Ecology and Evolution, University of Chicago}
\affil[c]{Department of Statistics, University of Michigan}
\affil[d]{Computer Science Division and Department of Statistics, University of California, Berkeley}
\affil[e]{Chan Zuckerberg Biohub, San Francisco} 
\affil[*]{To whom correspondence should be addressed: yss@berkeley.edu}

\date{}

\begin{document}

\maketitle

\begin{abstract}
    Studying how diverse human populations are related is of historical and anthropological interest, in addition to providing a realistic null model for testing for signatures of natural selection or disease associations.  Furthermore, understanding the demographic histories of other species is playing an increasingly important role in conservation genetics.  A number of statistical methods have been developed to infer population demographic histories using whole-genome sequence data, with recent advances focusing on allowing for more flexible modeling choices, scaling to larger data sets, and increasing statistical power.  Here we review coalescent hidden Markov models, a powerful class of population genetic inference methods that can effectively utilize linkage disequilibrium information.  We highlight recent advances, give advice for practitioners, point out potential pitfalls, and present possible future research directions.
\end{abstract}

\section{Introduction}
Using genetic data to understand the history of a population has been a long-standing goal of population genetics \cite{Cavalli1996}, and the emergence of massive data sets with individuals from many populations (e.g., \cite{Sudlow2015,Consortium2015,Mallick2016}), often including ancient samples \cite{Mathieson2015}, have enabled the inference of increasingly realistic models of the genetic history of human populations, e.g., \cite{Moorjani2013,Raghavan2015,Malaspinas2016}.  The progress in other species is no less impressive, with demographic models inferred for dogs \cite{vonHoldt2010}, horses, \cite{Warmuth2012}, pigs \cite{Frantz2015}, and many others.

These demographic models are frequently of interest in their own right for historical or anthropological reasons, and failing to account for demographic history when performing tests of neutrality \cite{Nielsen2005}, disease associations, \cite{Mathieson2012}, or recombination rate inference \cite{Johnston2012,Kamm2016} can lead to spurious results.  Demographic models also play an important role in conservation genetics, informing breeding strategies for maintaining genetic diversity in endangered populations, e.g., \cite{Mays2018}.

Yet, inferring complicated demographic models --- often including multiple populations with continuous migration, admixture events, and changes in effective population size --- is challenging both statistically and computationally, and numerous methods have been developed to address this problem.  Even under neutral evolution, computing the likelihood of observing a set of genotypes given a demographic model is computationally and analytically intractable.  Hence, demographic inference methods must make simplifying approximations and can be broadly divided into three classes: those based on allele frequencies; those based on identity-by-descent (IBD) or identity-by-state (IBS); and coalescent hidden Markov models (coalescent-HMMs).

Allele frequency-based methods summarize a collection of DNA sequence data as the multipopulation sample frequency spectrum (SFS) and use these summary statistics to infer either parametric \cite{Gutenkunst2009,Excoffier2013,Bhaskar2015,Jouganous2017,Kamm2018} or non-parametric \cite{Waltoft2017} models.  For computational purposes, these methods assume that all loci are independently evolving, an assumption obviously violated by physically-linked loci (although this has recently been relaxed to allow pairwise dependencies \cite{Ragsdale2017}).  This necessarily ignores the rich information contained in such linkage.  Yet, these methods tend to be very fast, with recent methods capable of scaling to data sets with hundreds of individuals from tens of populations \cite{Kamm2018}, making them ideal for quickly exploring a wide variety of potential models (e.g., testing models with different number of admixture events).  Yet, there are a number of concerns about statistical identifiability (\cite{Myers2008} but see \cite{Bhaskar2014}), power \cite{Terhorst2015,Baharian2018}, and stability \cite{Rosen2018}.

IBD- and IBS-based methods use patterns of pairwise haplotype sharing to infer demographic models, matching the distribution of observed IBD or IBS tract lengths to the distribution expected under the inferred demographic model.  While IBD-based methods, such as \cite{Palamara2012,Palamara2013,Browning2015}, can be powerful --- especially for learning about the recent past --- they rely on having access to unobserved IBD tracts.  Many methods have been developed for inferring IBD tracts \cite{Gusev2009,Browning2013}, but those methods rely either explicitly or implicitly on the unknown demographic history of the samples, resulting in a chicken/egg problem.  The effect of these assumptions on IBD-based methods has not been thoroughly explored, although see \cite{Tataru2014}.  To sidestep this issue, \cite{Harris2013} works directly with IBS tracts, a promising direction for further methodological development.

The final class of methods, coalescent-HMMs, is the focus of this review.  Below, we provide a historical overview of coalescent-HMMs and present a unifying framework for them.  We then explore recent advances in the field; discuss caveats, pitfalls, and best practices for applying coalescent-HMMs to data; and conclude with open problems and promising future research directions.

\section{A brief history of coalescent-HMMs}
All coalescent-HMMs can trace back to the seminal work of \cite{Wiuf1999}.  The coalescent --- a stochastic model of the genealogy of a sample of homologous chromosomes --- was first developed for a single non-recombining locus \cite{Kingman1982} and then extended to incorporate recombination \cite{Griffiths1996}, but had always been thought of as a process going backward in time.  In \cite{Wiuf1999} the multi-locus coalescent was viewed not as a process through time, but rather as a process along the genome (the so-called \emph{sequential coalescent}).  Unfortunately, the sequential coalescent was analytically complicated and non-Markovian (the genealogy at a locus depended on the genealogies at all previous loci).  Simpler Markovian models were later proposed (the \emph{sequentially Markovian coalescent}; SMC) \cite{McVean2005, Marjoram2006, Hobolth2014} that were highly accurate approximations of the original model \cite{Wilton2015}.  

Under the SMC, sequence data could be viewed as coming from a hidden Markov model (HMM) \cite{Rabiner1989} by treating the genealogy of the sampled individuals at a given locus as an unobserved, latent variable.  Because the demographic model impacts the distribution of genealogies (e.g., without migration, samples from different populations cannot have a common ancestor more recent than the divergence of those populations) and the observed sequence data are directly dependent on the underlying genealogy, coalescent-HMM methods have the potential to be extremely powerful.  An additional benefit of coalescent-HMMs is that the HMM framework enables integrating over all possible genealogies to make inferences about the demographic model --- even if there is substantial uncertainty about the genealogy of a given sample, the set of genealogies likely to have given rise to that sample may be highly informative about its demographic history.

In principle, the HMM framework enables efficient inference of demographic parameters, but there are a number of complications.  First, except for very special cases (e.g., Kalman Filters \cite{Kalman1960} and iHMMs \cite{Beal2001}), HMM algorithms require that the state space of the latent variable be finite; this is problematic in the coalescent-HMM case since the genealogy at a given locus has an uncountably infinite, continuous component (the lengths of the branches of the tree).  All coalescent-HMMs work around this issue by discretizing time.  Having a finite state space is not enough for efficient inference, however, as the number of tree topologies grows super-exponentially in the sample size, making the full coalescent-HMM impractical for all but the smallest sample sizes.  The menagerie of coalescent-HMM methods then arises by making different approximations to this idealized coalescent-HMM: instead of keeping track of the whole genealogy of the sample as a latent variable, these methods only track some subset of the genealogy or some of its features.

\begin{figure}
\centering
\includegraphics[width=0.8\textwidth]{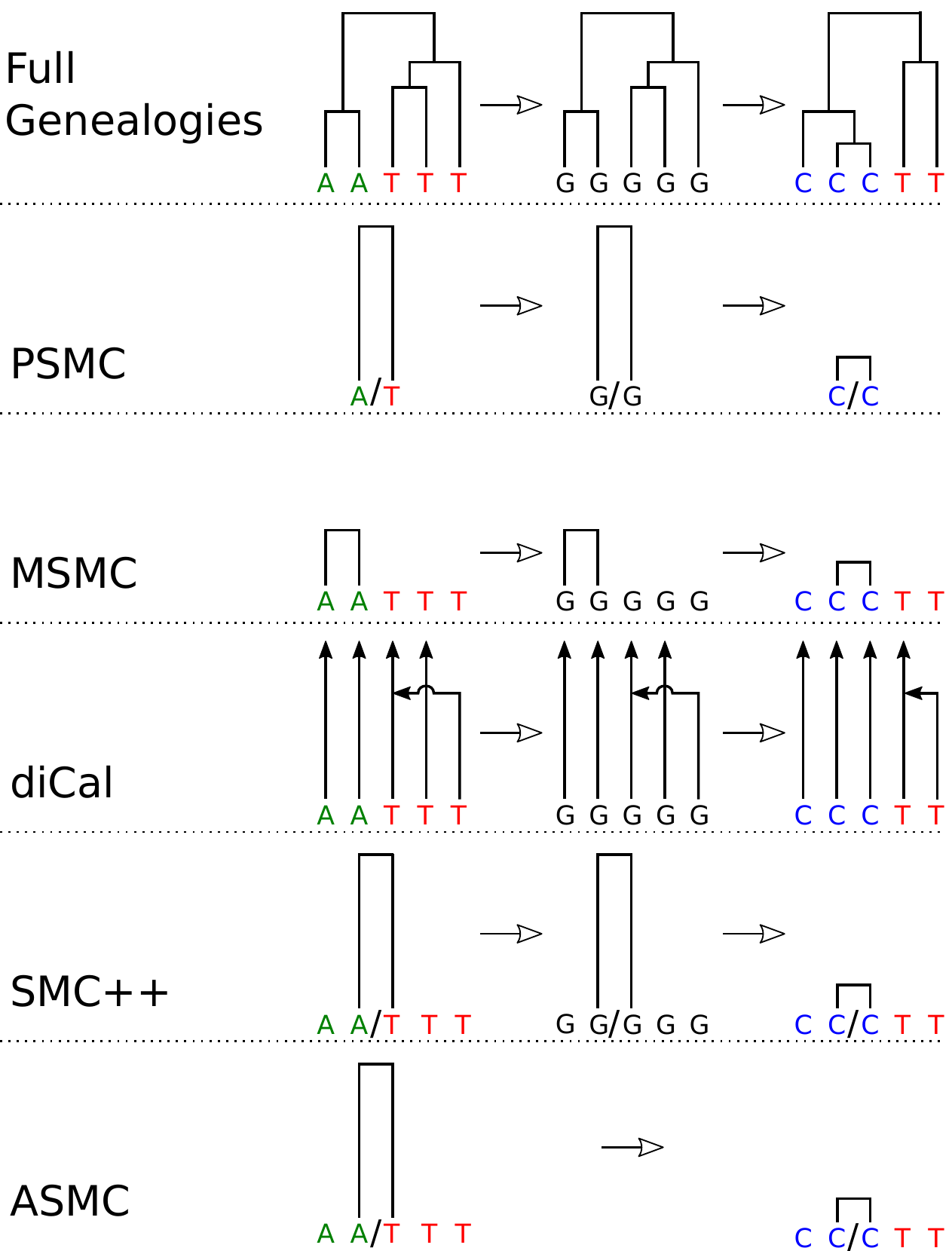}
\caption{The sequentially Markovian coalescent views the genealogy relating a sample of individuals as a sequence of trees along the genome.  The number of possible trees relating a sample grows super exponentially with sample size, making such a model computationally intractable for inference.  The commonly used coalescent-HMMs make various simplifications to this full process.  PSMC, \smcpp, and ASMC only track the genealogy of a ``distinguished'' pair of haplotypes.  PSMC ignores the rest of the sample, while \smcpp and ASMC use the other samples to inform the genealogy of the distinguished pair.  ASMC was designed to work on genotype array data and so skips over sites not included on the array (middle genealogy).  MSMC tracks only the most recent coalescence event in the whole sample, while diCal tracks the first coalescence event involving a particular haplotype.}
\label{fig:cartoon}
\end{figure}

Briefly, CoalHMM \cite{Dutheil2009,Mailund2012}, originally developed for studying the divergence of great apes, assumes that there is one sampled genome per species and tracks only the topology of the genealogy and the branch of the species tree in which the lineages coalesce and cannot scale to more than a few species.  PSMC \cite{Li2011} can be applied only to a pair of genomes but exactly tracks their genealogy up to the discretization of time and some simplifying approximations to the SMC.  The ideas underlying PSMC were extended to handle up to tens of genomes in MSMC \cite{Schiffels2014}, which tracks only the first coalescence time and which individuals were involved in the first coalescence event.  The first version of diCal \cite{Sheehan2013}, inspired by the copying model of \cite{Li2003} and the subsequent work on conditional sampling distributions (CSDs) \cite{Paul2010,Paul2011}, considers a particular haplotype and tracks when and with which other haplotype that haplotype first coalesces, with some approximations for computational efficiency.  PSMC makes the fewest simplifying assumptions, but as it can only be applied to two haplotypes it is less powerful than MSMC or diCal, especially in the recent past.

Furthermore, the different methods allow for the inference of different types of demographic models.  PSMC, MSMC, and diCal v1 can all infer piece-wise constant population size histories for a single panmictic population.  CoalHMM and MSMC are both capable of making inferences about multiple populations: CoalHMM fits a parametric model, directly inferring ancestral population sizes and divergence times between the populations, and MSMC performs non-parametric inference, reporting ``cross-coalescence rate'' curves (CCRs).  While such CCRs have been interpreted in terms of divergence times \cite{Schiffels2014,Mallick2016}, a thorough exploration of what types of models can give rise to a particular CCR has not been performed: if the goal of a study is to fit a particular demographic model (e.g., a two population isolation migration model), the CCR curves output by MSMC can be a useful diagnostic, but are difficult to interpret and cannot replace parametric model fitting.  All of the coalescent-HMMs discussed here are summarized visually in Figure~\ref{fig:cartoon}.

\section{Recent advances}
In response to many of the aforementioned issues, there has been much progress in coalescent-HMM methodology.  In particular, diCal version 2 allows for the parametric inference of more complex demographic models involving several populations, and \smcpp and ASMC push the boundaries of scalability for coalescent-HMMs.

Building on the first version of diCal \cite{Sheehan2013} and advances to the CSD framework \cite{Davison2009,Steinrucken2013}, diCal v2 \cite{Steinrucken2015} was developed to perform parametric inference of essentially arbitrarily complex demographic models, including estimating divergence times, continuous and pulse migration, and population sizes with possible exponential growth.  The method can scale to tens of haplotypes and has been tested on models with three populations, but can, in principal, handle an arbitrary number of populations (but at an increased computational cost). Like diCal v1, version 2 also considers a particular haplotype, and keeps track of when and with which other haplotype it first coalesces.  Since first coalescence events tend to happen in the recent past, this makes diCal well-powered to investigate recent history, such as the peopling of the Americas \cite{Raghavan2015,Moreno2018}.  It is also possible to use coalescent-HMMs in a hypothesis testing framework: diCal v2 was used in \cite[Supplementary Information, section 18.4]{Moreno2018} to test a null model of a clean split between two populations against a model of gene flow following that split, providing a principled and powerful technique for performing model selection and also for falsifying testable hypotheses.  Furthermore, the CSD framework used by  diCal v2 allows it to perform local ancestry or admixture calling, which was recently used to infer tracts of Neanderthal introgression in modern humans \cite{Steinrucken2018}.

\smcpp \cite{Terhorst2016} combines the power of SFS-based methods with the simplicity of PSMC and its lack of making assumptions beyond the SMC.  \smcpp tracks the coalescence time of a single ``distinguished'' pair of lineages like PSMC, but then computes the likelihood of observing the sequence data of both the distinguished lineages and the rest of the sample.  Like PSMC, \smcpp does not require phased data.  The simplicity of the hidden state allows \smcpp to scale to sample sizes in the hundreds, which is about an order of magnitude larger than any other coalescent-HMM presented above.  This in turn gives \smcpp substantial power in both the recent and ancient past.  It also achieves a substantial speedup by taking advantage of the fact that genotype data contains long stretches of non-segregating loci which may be effectively ``skipped over'' --- an idea similar in spirit to \cite{Paul2012}. Furthermore, instead of inferring unrealistic piece-wise constant population sizes, \smcpp fits population sizes as smooth splines, reflecting a more realistic scenario of non-instantaneous population size changes.  \smcpp is also capable of inferring divergence times for a pair of populations but currently makes the assumption that there has been no migration between the populations since their divergence, which may not be appropriate for some populations.

Recently, ASMC \cite{Palamara2018} has been developed to extend \smcpp to genotype array data by accounting for ascertainment bias in the frequency of alleles measured by genotyping arrays.  ASMC also takes advantage of certain symmetries in computing likelihoods in the underlying HMM to obtain extremely fast runtimes --- an idea first explored in \cite{Harris2014}.  In fact, its speed allowed ASMC to be run on all pairs of haplotypes from 113,756 phased British individuals \cite{Sudlow2015} although still at considerable computational cost.

\begin{figure}
\centering
\includegraphics[width=.8\textwidth]{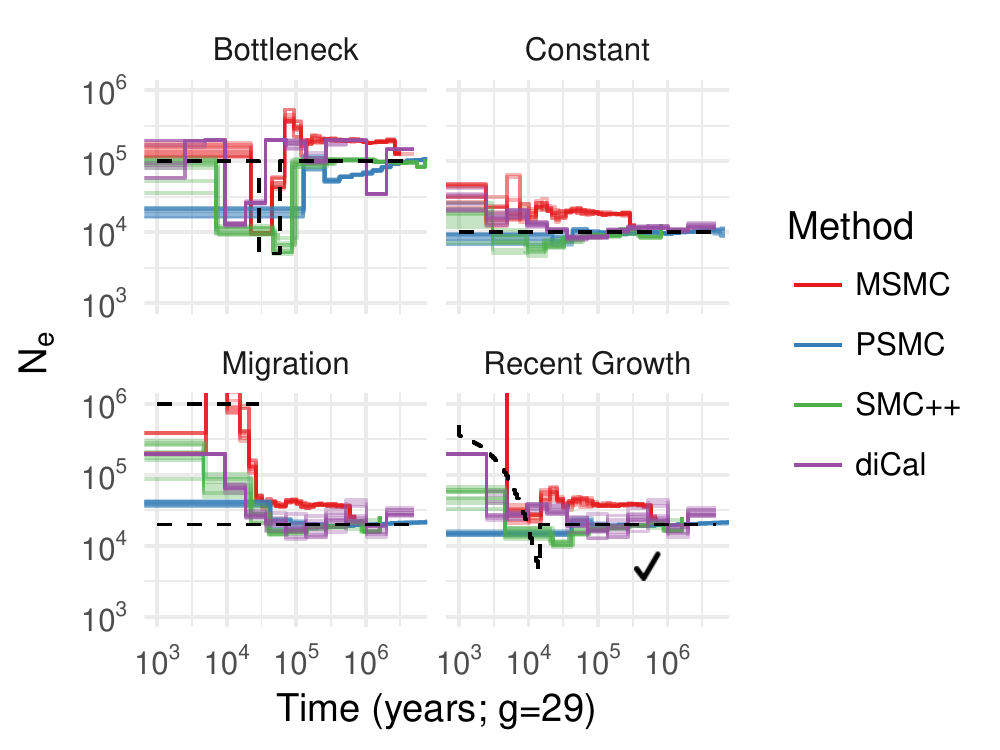}
    \caption{Performance of various coalescent-HMMs on simulated data.  The scenarios considered here are: a population experiencing a sharp bottleneck; a single panmictic population of constant size; samples from a large population that is exchanging migrants with a smaller population; and a population that has recently experienced exponential growth.  Each scenario has 10 replicate data sets, with each data set containing 30 haploids with eight 125~Mb chromosomes per haploid.  PSMC was run with the options `-N 25 -p 4+20*3+4' on a single pair of haploids. MSMC was run with the default hyperparameter settings with the `fixedRecombination' flag, using only 4 of the 30 haploids.  The same 4 haploids were used for diCal v2, and inference was performed by taking the composite likelihood over all pairs of those 4 haplotypes, and running 30 EM iterations.  \smcpp was run with the `--timepoints 33' and `--thinning 500' options.}
    \label{fig:sim}
\end{figure}

We present the results of a small simulation study in Figure~\ref{fig:sim} showing the performance of various coalescent-HMMs for a number of common demographic scenarios. The four scenarios considered were:
\begin{itemize}
    \item A sharp bottleneck.
    \item Constant size ($N_e=10^4$).
    \item An isolation-with-migration model involving two populations.
    \item Exponential growth beginning 500 generations ago.
\end{itemize}
For each scenario, we used \texttt{msprime} \cite{Kelleher2016} to simulate 10 replicate data sets each consisting of 30 haploids with eight 125~Mb chromosomes per haploid.  The code used to simulate data and infer population sizes is fully reproducible and available at https://github.com/terhorst/coal\_hmm\_review.

\section{Caveats, pitfalls, and best practices}
Despite their power and flexibility, coalescent-HMMs are not without their pitfalls.  All coalescent-HMMs contain tuning parameters that are crucial for good performance.  A critical factor is the way that time is discretized.  Finer discretization leads to a more accurate approximation, but the runtime of all methods is impacted by the number of discretization points so care is needed to balance computational and accuracy considerations.  Additionally, all of the methods discussed above, save \smcpp, group adjacent loci and assume that they have the same genealogy.  This assumption helps to substantially decrease the runtime, but is certainly violated in practice.  Depending on the method and application, it may be acceptable to perform the grouping at a kb scale, but care should be taken to check that such grouping does not influence the results.  Furthermore, the likelihoods optimized by coalescent-HMMs--and demographic inference methods more broadly--tend to be highly non-convex: they have many local optima and so initializing the methods at different initial values will result in different inferred models.  As such it is important to take the best of several runs with different initializations as the final inferred model.

Users should also be careful about the model used to fit data.  As an example, \smcpp infers population splits in the absence of gene flow.  If there has been pervasive migration between the populations of interest, then the model inferred by \smcpp is unlikely to be reflective of reality.  Also, note that even seemingly non-parametric methods, like PSMC, make implicit assumptions such as the data coming from a single panmictic population that has been evolving neutrally.  Recent studies \cite{Hawks2017,Schrider2016} used simulated data to investigate these model violations and showed that pervasive selective sweeps or structured populations bias the results of coalescent-HMMs.  Another study \cite{Beichman2017} showed that when applied to simulated data, coalescent-HMMs infer models that have an expected SFS similar to that of the data, but when applied to real data the SFS of the inferred models does not match that of the data.  This suggests that real data violate the idealized models that are commonly used for simulation and inference.

We also urge caution in over-interpreting the results of any demographic inference method.  For instance, all methods infer ``effective population sizes'', which are defined as the inverse of the coalescence rate for a pair of haplotypes.  Under many models effective population size is correlated with census population size, but does not need to be; e.g., a structured population will have a larger effective size than a panmictic population of the same census size.

To avoid the aforementioned pitfalls, we recommend using multiple methods utilizing different aspects of the data, such as frequency-based methods \emph{and} coalescent-HMMs.  While the exact models inferred by different methods will differ, one can have some confidence in aspects of the model that are robustly inferred across methods.  We also recommend using the results of either a pilot run of the coalescent-HMM or the results of another method (or even PCA \cite{Price2006,Novembre2008}, or STRUCTURE-like programs \cite{Pritchard2000,Alexander2009,Raj2014,Cabreros2017}) to inform model selection --- e.g., if the data appear to come from unadmixed populations based on this initial fit, it may be appropriate to assume a clean split model instead of modeling gene flow.  After fitting a model, it is crucial to measure goodness-of-fit, for example by comparing the SFS and MSMC's CCR curves for data simulated from the inferred models to those computed directly from the real data.

It is also important to understand sources of bias and noise present in data.  Because most coalescent-HMMs make use of both segregating and non-segregating sites it is crucial to use ``masks'' that indicate which regions of the genome have been reliably genotyped.  Additionally, when working with ancient DNA, which tends to show an excess of transitions due to postmortem cytosine deamination \cite{Dabney2013}, we have found that restricting analysis to only transversions and adjusting the mutation rate correspondingly improves inference.

Finally, as with any statistical analysis, it is important to study uncertainty in the inferred model, e.g., by bootstrapping, either parametric via simulation or non-parametric by resampling the data as in \cite{Li2011}.  While parametric bootstrapping is more straightforward, it is only capable of estimating uncertainty in the estimation procedure, whereas non-parametric bootstrapping captures uncertainty in both modeling and estimation, but cannot reveal bias in the estimates.  Note that in  demographic inference, bootstrapping does not produce statistically valid confidence intervals due to using the data to perform model selection prior to estimating statistical uncertainty, but providing some quantification of uncertainty is still important.

\section{Outlook}
While there has been much recent work on improving the flexibility, and computational and statistical efficiency of coalescent-HMMs, there are still a number of open problems and interesting directions for future research.

As alluded to above, when the sample size is greater than 2, every coalescent-HMM treats only a part of the whole genealogy relating the sample as a latent variable.  Such choices are made primarily for analytic convenience to ensure computational tractability, based on intuition.  Tree length has recently been explored as such a choice \cite{Miroshnikov2017}.  Finding more optimal ways of modeling genealogical information using a small number of discrete parameters remains a challenging open problem.

While coalescent-HMMs work extremely well on simulated data, they, like most inference methods in population genetics, seem to be less stable on real data \cite{Beichman2017}.  This is likely due to rampant model misspecifications: coalescent-HMMs make many unrealistic assumptions, such as assuming constant recombination \cite{Myers2005,Kong2010} and mutation \cite{Kong2012,Jonsson2017,Smith2018} rates across the genome.  In addition, all methods must simplify the ``true'' demographic model: reality is always more complicated than any model with a handful of parameters.  The impact of these misspecifications has not been thoroughly explored, and there is a need for more robust methods.

A major challenge, especially in studying non-model organisms, is that with the exception of PSMC and \smcpp, coalescent-HMMs are currently unable to handle unphased data.  Overcoming this challenge is an important task for future methods.

Lastly, despite their excellent behavior in practice, our understanding of coalescent-HMMs is based entirely on intuition and numerical experiments.  In contrast to frequency-based methods, which have a rich literature on their theoretical properties \cite{Myers2008,Bhaskar2014,Terhorst2015,Baharian2018,Rosen2018}, coalescent-HMMs are poorly understood from a theoretical perspective.  While there has been some work on how accurately demographic history can be inferred directly from genealogies \cite{Kim2015,Johndrow2018}, in the more realistic coalescent-HMM setting even the basic question of whether demographic models are statistically identifiable is unanswered.

\section*{Acknowledgments}
This work is supported in part by an NIH grant R01-GM094402, and a Packard Fellowship for Science and Engineering. Y.S.S. is a Chan Zuckerberg Biohub investigator.

\bibliographystyle{my_unsrt}  
\bibliography{bib}

\begin{thebibliography}{10}

\bibitem{Cavalli1996}
L.~Luca Cavalli-Sforza, Paolo Menozzi, and Alberto Piazza.
\newblock {\em The History and Geography of Human Genes}.
\newblock Princeton paperbacks. Princeton University Press, 1996.
\bibitem{Sudlow2015}
Cathie Sudlow, John Gallacher, Naomi Allen, Valerie Beral, Paul Burton, John
  Danesh, Paul Downey, Paul Elliott, Jane Green, Martin Landray, et~al.
\newblock {UK B}iobank: {A}n open access resource for identifying the causes of
  a wide range of complex diseases of middle and old age.
\newblock {\em PLOS Medicine}, 12(3):1--10, 2015.
\bibitem{Consortium2015}
The 1000 Genomes~Project Consortium.
\newblock A global reference for human genetic variation.
\newblock {\em Nature}, 526:68--74, 2015.
\bibitem{Mallick2016}
Swapan Mallick, Heng Li, Mark Lipson, Iain Mathieson, Melissa Gymrek, Fernando
  Racimo, Mengyao Zhao, Niru Chennagiri, Susanne Nordenfelt, Arti Tandon,
  et~al.
\newblock The {S}imons {G}enome {D}iversity {P}roject: 300 genomes from 142
  diverse populations.
\newblock {\em Nature}, 538(7624):201--206, 2016.
\begin{quotation}\noindent** (of outstanding interest): One of the most
  extensive and diverse set of human genomes to date. Includes samples from 142
  populations located throughout the world.\end{quotation}

\bibitem{Mathieson2015}
Iain Mathieson, Iosif Lazaridis, Nadin Rohland, Swapan Mallick, Nick Patterson,
  Song{\"u}l~Alpaslan Roodenberg, Eadaoin Harney, Kristin Stewardson, Daniel
  Fernandes, Mario Novak, et~al.
\newblock Genome-wide patterns of selection in 230 ancient {E}urasians.
\newblock {\em Nature}, 528(7583):499--503, 2015.
\bibitem{Moorjani2013}
Priya Moorjani, Kumarasamy Thangaraj, Nick Patterson, Mark Lipson, Po-Ru Loh,
  Periyasamy Govindaraj, Bonnie Berger, David Reich, and Lalji Singh.
\newblock Genetic evidence for recent population mixture in {I}ndia.
\newblock {\em American Journal of Human Genetics}, 93(3):422--438, 2013.
\bibitem{Raghavan2015}
Maanasa Raghavan, Matthias Steinr{\"u}cken, Kelley Harris, Stephan Schiffels,
  Simon Rasmussen, Michael DeGiorgio, Anders Albrechtsen, Cristina Valdiosera,
  Mar{\'\i}a~C. {\'A}vila-Arcos, Anna-Sapfo Malaspinas, et~al.
\newblock Genomic evidence for the {P}leistocene and recent population history
  of {N}ative {A}mericans.
\newblock {\em Science}, 349(6250), 2015.
\bibitem{Malaspinas2016}
Anna-Sapfo Malaspinas, Michael~C. Westaway, Craig Muller, Vitor~C. Sousa, Oscar
  Lao, Isabel Alves, Anders Bergstr{\"o}m, Georgios Athanasiadis, Jade~Y.
  Cheng, Jacob~E. Crawford, et~al.
\newblock A genomic history of {A}boriginal {A}ustralia.
\newblock {\em Nature}, 538:207--214, 2016.
\bibitem{vonHoldt2010}
Bridgett~M. vonHoldt, John~P. Pollinger, Kirk~E. Lohmueller, Eunjung Han,
  Heidi~G. Parker, Pascale Quignon, Jeremiah~D. Degenhardt, Adam~R. Boyko,
  Dent~A. Earl, Adam Auton, et~al.
\newblock Genome-wide {SNP} and haplotype analyses reveal a rich history
  underlying dog domestication.
\newblock {\em Nature}, 464:898--902, 2010.
\bibitem{Warmuth2012}
Vera Warmuth, Anders Eriksson, Mim~Ann Bower, Graeme Barker, Elizabeth Barrett,
  Bryan~Kent Hanks, Shuicheng Li, David Lomitashvili, Maria Ochir-Goryaeva,
  Grigory~V. Sizonov, et~al.
\newblock Reconstructing the origin and spread of horse domestication in the
  {E}urasian steppe.
\newblock {\em Proceedings of the National Academy of Sciences},
  109(21):8202--8206, 2012.
\bibitem{Frantz2015}
Laurent A.~F. Frantz, Joshua~G. Schraiber, Ole Madsen, Hendrik-Jan Megens, Alex
  Cagan, Mirte Bosse, Yogesh Paudel, Richard P. M.~A. Crooijmans, Greger
  Larson, and Martien A.~M. Groenen.
\newblock Evidence of long-term gene flow and selection during domestication
  from analyses of {E}urasian wild and domestic pig genomes.
\newblock {\em Nature Genetics}, 47:1141--1148, 2015.
\bibitem{Nielsen2005}
Rasmus Nielsen, Scott Williamson, Yuseob Kim, Melissa~J. Hubisz, Andrew~G.
  Clark, and Carlos Bustamante.
\newblock Genomic scans for selective sweeps using {SNP} data.
\newblock {\em Genome Research}, 15(11):1566--1575, 2005.
\bibitem{Mathieson2012}
Iain Mathieson and Gil McVean.
\newblock Differential confounding of rare and common variants in spatially
  structured populations.
\newblock {\em Nature Genetics}, 44:243--246, 2012.
\bibitem{Johnston2012}
Henry~R. Johnston and David~J. Cutler.
\newblock Population demographic history can cause the appearance of
  recombination hotspots.
\newblock {\em The American Journal of Human Genetics}, 90(5):774--783, 2012.
\bibitem{Kamm2016}
John~A. Kamm, Jeffrey~P. Spence, Jeffrey Chan, and Yun~S. Song.
\newblock Two-locus likelihoods under variable population size and fine-scale
  recombination rate estimation.
\newblock {\em Genetics}, 203(3):1381--1399, 2016.
\bibitem{Mays2018}
Herman~L. Mays~Jr., Chih-Ming Hung, Pei-Jen Shaner, James Denvir, Megan
  Justice, Shang-Fang Yang, Terri~L. Roth, David~A. Oehler, Jun Fan, Swanthana
  Rekulapally, and Donald~A. Primerano.
\newblock Genomic analysis of demographic history and ecological niche modeling
  in the endangered {S}umatran rhinoceros \emph{Dicerorhinus sumatrensis}.
\newblock {\em Current Biology}, 28(1):70--76.e4, 2018.
\bibitem{Gutenkunst2009}
Ryan~N. Gutenkunst, Ryan~D. Hernandez, Scott~H. Williamson, and Carlos~D.
  Bustamante.
\newblock Inferring the joint demographic history of multiple populations from
  multidimensional {SNP} frequency data.
\newblock {\em PLOS Genetics}, 5(10):e1000695, 2009.
\bibitem{Excoffier2013}
Laurent Excoffier, Isabelle Dupanloup, Emilia Huerta-S{\'a}nchez, Vitor~C.
  Sousa, and Matthieu Foll.
\newblock Robust demographic inference from genomic and {SNP} data.
\newblock {\em PLOS Genetics}, 9(10):1--17, 2013.
\bibitem{Bhaskar2015}
Anand Bhaskar, Y.X.~Rachel Wang, and Yun~S. Song.
\newblock Efficient inference of population size histories and locus-specific
  mutation rates from large-sample genomic variation data.
\newblock {\em Genome Research}, 25(2):268--279, 2015.
\bibitem{Jouganous2017}
Julien Jouganous, Will Long, Aaron~P. Ragsdale, and Simon Gravel.
\newblock Inferring the joint demographic history of multiple populations:
  Beyond the diffusion approximation.
\newblock {\em Genetics}, 206(3):1549--1567, 2017.
\begin{quotation}\noindent* (of special interst): Uses a sparse approximation
  to Wright-Fisher dynamics to efficiently compute the SFS for multiple
  populations allowing for possible selection.\end{quotation}

\bibitem{Kamm2018}
John~A. Kamm, Jonathan Terhorst, Richard Durbin, and Yun~S. Song.
\newblock Efficiently inferring the demographic history of many populations
  with allele count data.
\newblock {\em bioRxiv}, 2018.
\newblock https://doi.org/10.1101/287268.
\begin{quotation}\noindent* (of special interest): Presents an extremely
  efficient method to compute the expected frequency spectrum of many
  populations, extending the applicability of frequency-based methods sample
  sizes in the hundreds for tens of populations.\end{quotation}

\bibitem{Waltoft2017}
Berit~Lindum Waltoft and Asger Hobolth.
\newblock Non-parametric estimation of population size changes from the site
  frequency spectrum.
\newblock {\em Stat Appl Genet Mol Biol}, 17(3), 2018.
\bibitem{Ragsdale2017}
Aaron~P. Ragsdale and Ryan~N. Gutenkunst.
\newblock Inferring demographic history using two-locus statistics.
\newblock {\em Genetics}, 206(2):1037--1048, 2017.
\bibitem{Myers2008}
Simon Myers, Charles Fefferman, and Nick Patterson.
\newblock Can one learn history from the allelic spectrum?
\newblock {\em Theoretical Population Biology}, 73(3):342--348, 2008.
\bibitem{Bhaskar2014}
Anand Bhaskar and Yun~S. Song.
\newblock Descartes' rule of signs and the identifiability of population
  demographic models from genomic variation data.
\newblock {\em Annals of Statistics}, 42(6):2469--2493, 2014.
\bibitem{Terhorst2015}
Jonathan Terhorst and Yun~S. Song.
\newblock Fundamental limits on the accuracy of demographic inference based on
  the sample frequency spectrum.
\newblock {\em Proceedings of the National Academy of Sciences},
  112(25):7677--7682, 2015.
\bibitem{Baharian2018}
Soheil Baharian and Simon Gravel.
\newblock On the decidability of population size histories from finite allele
  frequency spectra.
\newblock {\em Theoretical Population Biology}, 120:42--51, 2018.
\begin{quotation}\noindent* (of special interest): presents classes of
  piece-wise constant population size histories that are qualitatively and
  quantitatively dissimilar but produce provably similar frequency
  spectra.\end{quotation}

\bibitem{Rosen2018}
Zvi Rosen, Anand Bhaskar, Sebastien Roch, and Yun~S. Song.
\newblock Geometry of the sample frequency spectrum and the perils of
  demographic inference.
\newblock {\em bioRxiv}, 2017.
\newblock https://doi.org/10.1101/233908.
\bibitem{Palamara2012}
Pier~Francesco Palamara, Todd Lencz, Ariel Darvasi, and Itsik Pe'er.
\newblock Length distributions of identity by descent reveal fine-scale
  demographic history.
\newblock {\em The American Journal of Human Genetics}, 91(5):809--822, 2012.
\bibitem{Palamara2013}
Pier~Francesco Palamara and Itsik Pe'er.
\newblock Inference of historical migration rates via haplotype sharing.
\newblock {\em Bioinformatics}, 29(13):i180--i188, 2013.
\bibitem{Browning2015}
Sharon~R. Browning and Brian~L. Browning.
\newblock Accurate non-parametric estimation of recent effective population
  size from segments of identity by descent.
\newblock {\em The American Journal of Human Genetics}, 97(3):404--418, 2015.
\bibitem{Gusev2009}
Alexander Gusev, Jennifer~K. Lowe, Markus Stoffel, Mark~J. Daly, David
  Altshuler, Jan~L. Breslow, Jeffrey~M. Friedman, and Itsik Pe'er.
\newblock Whole population, genome-wide mapping of hidden relatedness.
\newblock {\em Genome Research}, 19(2):318--326, 2009.
\bibitem{Browning2013}
Brian~L. Browning and Sharon~R. Browning.
\newblock Detecting identity by descent and estimating genotype error rates in
  sequence data.
\newblock {\em The American Journal of Human Genetics}, 93(5):840--851, 2013.
\bibitem{Tataru2014}
Paula Tataru, Jasmine~A. Nirody, and Yun~S. Song.
\newblock {diCal-IBD:} demography-aware inference of identity-by-descent tracts
  in unrelated individuals.
\newblock {\em Bioinformatics}, 30(23):3430--3431, 2014.
\bibitem{Harris2013}
Kelley Harris and Rasmus Nielsen.
\newblock Inferring demographic history from a spectrum of shared haplotype
  lengths.
\newblock {\em PLOS Genetics}, 9(6):1--20, 2013.
\bibitem{Wiuf1999}
Carsten Wiuf and Jotun Hein.
\newblock Recombination as a point process along sequences.
\newblock {\em Theoretical Population Biology}, 55(3):248--259, 1999.
\bibitem{Kingman1982}
John F.~C. Kingman.
\newblock The coalescent.
\newblock {\em Stochastic Processes and their Applications}, 13(3):235--248,
  1982.
\bibitem{Griffiths1996}
Robert~C. Griffiths and Paul Marjoram.
\newblock Ancestral inference from samples of {DNA} sequences with
  recombination.
\newblock {\em Journal of Computational Biology}, 3(4):479--502, 1996.
\newblock PMID: 9018600.
\bibitem{McVean2005}
Gilean~A.T. McVean and Niall~J. Cardin.
\newblock Approximating the coalescent with recombination.
\newblock {\em Philosophical Transactions of the Royal Society London B:
  Biological Sciences}, 360:1387--93, 2005.
\bibitem{Marjoram2006}
Paul Marjoram and Jeff~D. Wall.
\newblock Fast ``coalescent'' simulation.
\newblock {\em BMC Genetics}, 7(1):16, 2006.
\bibitem{Hobolth2014}
Asger Hobolth and Jens~Ledet Jensen.
\newblock Markovian approximation to the finite loci coalescent with
  recombination along multiple sequences.
\newblock {\em Theoretical Population Biology}, 98:48--58, 2014.
\bibitem{Wilton2015}
Peter~R. Wilton, Shai Carmi, and Asger Hobolth.
\newblock The {SMC}' is a highly accurate approximation to the ancestral
  recombination graph.
\newblock {\em Genetics}, 200(1):343--355, 2015.
\bibitem{Rabiner1989}
Lawrence~R. Rabiner.
\newblock A tutorial on hidden {M}arkov models and selected applications in
  speech recognition.
\newblock {\em Proceedings of the IEEE}, 77(2):257--286, 1989.
\bibitem{Kalman1960}
Rudolph~Emil Kalman.
\newblock A new approach to linear filtering and prediction problems.
\newblock {\em Transactions of the ASME--Journal of Basic Engineering},
  82(Series D):35--45, 1960.
\bibitem{Beal2001}
Matthew~J. Beal, Zoubin Ghahramani, and Carl~E. Rasmussen.
\newblock The infinite hidden {M}arkov model.
\newblock In T.~G. Dietterich, S.~Becker, and Z.~Ghahramani, editors, {\em
  Advances in Neural Information Processing Systems 14}, pages 577--584. MIT
  Press, 2002.
\bibitem{Dutheil2009}
Julien~Y. Dutheil, Ganesh Ganapathy, Asger Hobolth, Thomas Mailund, Marcy~K.
  Uyenoyama, and Mikkel~H. Schierup.
\newblock Ancestral population genomics: The coalescent hidden {M}arkov model
  approach.
\newblock {\em Genetics}, 183(1):259--274, 2009.
\bibitem{Mailund2012}
Thomas Mailund, Anders~E. Halager, and Michael Westergaard.
\newblock Using colored petri nets to construct coalescent hidden {M}arkov
  models: Automatic translation from demographic specifications to efficient
  inference methods.
\newblock In Serge Haddad and Lucia Pomello, editors, {\em Application and
  Theory of Petri Nets}, pages 32--50, Berlin, Heidelberg, 2012. Springer
  Berlin Heidelberg.
\bibitem{Li2011}
Heng Li and Richard Durbin.
\newblock Inference of human population history from individual whole-genome
  sequences.
\newblock {\em Nature}, 475:493--496, 2011.
\bibitem{Schiffels2014}
Stephan Schiffels and Richard Durbin.
\newblock Inferring human population size and separation history from multiple
  genome sequences.
\newblock {\em Nature Genetics}, 46:919--925, 2014.
\bibitem{Sheehan2013}
Sara Sheehan, Kelley Harris, and Yun~S. Song.
\newblock Estimating variable effective population sizes from multiple genomes:
  A sequentially {M}arkov conditional sampling distribution approach.
\newblock {\em Genetics}, 194(3):647--662, 2013.
\bibitem{Li2003}
Na~Li and Matthew Stephens.
\newblock Modeling linkage disequilibrium and identifying recombination
  hotspots using single-nucleotide polymorphism data.
\newblock {\em Genetics}, 165(4):2213--2233, 2003.
\bibitem{Paul2010}
Joshua~S. Paul and Yun~S. Song.
\newblock A principled approach to deriving approximate conditional sampling
  distributions in population genetics models with recombination.
\newblock {\em Genetics}, 186(1):321--338, 2010.
\bibitem{Paul2011}
Joshua~S. Paul, Matthias Steinr{\"u}cken, and Yun~S. Song.
\newblock An accurate sequentially {M}arkov conditional sampling distribution
  for the coalescent with recombination.
\newblock {\em Genetics}, 187(4):1115--1128, 2011.
\bibitem{Davison2009}
Dan Davison, Jonathan~K. Pritchard, and Graham Coop.
\newblock An approximate likelihood for genetic data under a model with
  recombination and population splitting.
\newblock {\em Theoretical Population Biology}, 75(4):331--345, 2009.
\bibitem{Steinrucken2013}
Matthias Steinr{\"u}cken, Joshua~S. Paul, and Yun~S. Song.
\newblock A sequentially {M}arkov conditional sampling distribution for
  structured populations with migration and recombination.
\newblock {\em Theoretical Population Biology}, 87:51--61, 2013.
\bibitem{Steinrucken2015}
Matthias Steinr{\"u}cken, John~A. Kamm, and Yun~S. Song.
\newblock Inference of complex population histories using whole-genome
  sequences from multiple populations.
\newblock {\em bioRxiv}, 2015.
\newblock http://dx.doi.org/10.1101/026591.
\bibitem{Moreno2018}
J.~V{\'\i}ctor Moreno-Mayar, Ben~A. Potter, Lasse Vinner, Matthias
  Steinr{\"u}cken, Simon Rasmussen, Jonathan Terhorst, John~A. Kamm, Anders
  Albrechtsen, Anna-Sapfo Malaspinas, Martin Sikora, et~al.
\newblock Terminal {P}leistocene {A}laskan genome reveals first founding
  population of {N}ative {A}mericans.
\newblock {\em Nature}, 553:203--207, 2018.
\begin{quotation}\noindent* (of special interest): Studies the peopling of the
  Americas, making use of ancient genomes and combining frequency-based and
  coalescent-HMM methods for robust demographic inference.\end{quotation}

\bibitem{Steinrucken2018}
Matthias Steinr{\"u}cken, Jeffrey~P. Spence, John A. John~A. Kamm, Emilia
  Wieczorek, and Yun~S. Song.
\newblock Model-based detection and analysis of introgressed {N}eanderthal
  ancestry in modern humans.
\newblock {\em Molecular Ecology}, 2018.
\bibitem{Terhorst2016}
Jonathan Terhorst, John~A. Kamm, and Yun~S. Song.
\newblock Robust and scalable inference of population history from hundreds of
  unphased whole genomes.
\newblock {\em Nature Genetics}, 49:303--309, 2017.
\begin{quotation}\noindent** (of outstanding interest): Presents a
  coalescent-HMM that essentially combines PSMC with frequency-based methods
  for a powerful, yet scalable tool for demographic inference.\end{quotation}

\bibitem{Paul2012}
Joshua~S. Paul and Yun~S. Song.
\newblock Blockwise {HMM} computation for large-scale population genomic
  inference.
\newblock {\em Bioinformatics}, 28(15):2008--2015, 2012.
\bibitem{Palamara2018}
Pier~Francesco Palamara, Jonathan Terhorst, Yun~S. Song, and Alkes~L. Price.
\newblock High-throughput inference of pairwise coalescence times identifies
  signals of selection and enriched disease heritability.
\newblock {\em Nature Genetics}, 2018.
\newblock In press.
\begin{quotation}\noindent* (of special interest): Extends ideas from
  SMC\texttt{++} to data from genotype array data sets and presents the
  largest-scale application of coalescent-HMMs to date.\end{quotation}

\bibitem{Harris2014}
Kelley Harris, Sara Sheehan, John~A. Kamm, and Yun~S. Song.
\newblock Decoding coalescent hidden {M}arkov models in linear time.
\newblock In Roded Sharan, editor, {\em Research in Computational Molecular
  Biology}, pages 100--114, Cham, 2014. Springer International Publishing.
\bibitem{Kelleher2016}
Jerome Kelleher, Alison~M Etheridge, and Gilean McVean.
\newblock Efficient coalescent simulation and genealogical analysis for large
  sample sizes.
\newblock {\em PLOS Computational Biology}, 12(5):e1004842, 2016.
\begin{quotation}\noindent** (of outstanding interest): Presents
  \texttt{msprime}: simulation software capable of simulating data under the
  full coalescent with recombination orders of magnitude faster than other
  simulators.\end{quotation}

\bibitem{Hawks2017}
John Hawks.
\newblock Introgression makes waves in inferred histories of effective
  population size.
\newblock {\em Human Biology}, 89(1):67--80, 2017.
\bibitem{Schrider2016}
Daniel~R. Schrider, Alexander~G. Shanku, and Andrew~D. Kern.
\newblock Effects of linked selective sweeps on demographic inference and model
  selection.
\newblock {\em Genetics}, 204(3):1207--1223, 2016.
\bibitem{Beichman2017}
Annabel~C. Beichman, Tanya~N. Phung, and Kirk~E. Lohmueller.
\newblock Comparison of single genome and allele frequency data reveals
  discordant demographic histories.
\newblock {\em G3: Genes, Genomes, Genetics}, 7(11):3605--3620, 2017.
\bibitem{Price2006}
Alkes~L Price, Nick~J Patterson, Robert~M Plenge, Michael~E Weinblatt, Nancy~A
  Shadick, and David Reich.
\newblock Principal components analysis corrects for stratification in
  genome-wide association studies.
\newblock {\em Nature Genetics}, 38:904--909, 2006.
\bibitem{Novembre2008}
John Novembre, Toby Johnson, Katarzyna Bryc, Zolt{\'a}n Kutalik, Adam~R. Boyko,
  Adam Auton, Amit Indap, Karen~S. King, Sven Bergmann, Matthew~R. Nelson,
  et~al.
\newblock Genes mirror geography within {E}urope.
\newblock {\em Nature}, 456:98--101, 2008.
\bibitem{Pritchard2000}
Jonathan~K. Pritchard, Matthew Stephens, and Peter Donnelly.
\newblock Inference of population structure using multilocus genotype data.
\newblock {\em Genetics}, 155(2):945--959, 2000.
\bibitem{Alexander2009}
David~H. Alexander, John Novembre, and Kenneth Lange.
\newblock Fast model-based estimation of ancestry in unrelated individuals.
\newblock {\em Genome Research}, 2009.
\bibitem{Raj2014}
Anil Raj, Matthew Stephens, and Jonathan~K. Pritchard.
\newblock fast{STRUCTURE}: Variational inference of population structure in
  large {SNP} data sets.
\newblock {\em Genetics}, 197(2):573--589, 2014.
\bibitem{Cabreros2017}
Irineo Cabreros and John~D. Storey.
\newblock A nonparametric estimator of population structure unifying admixture
  models and principal components analysis.
\newblock {\em bioRxiv}, 2017.
\newblock https://doi.org/10.1101/240812.
\bibitem{Dabney2013}
Jesse Dabney, Matthias Meyer, and Svante P{\"a}{\"a}bo.
\newblock Ancient {DNA} damage.
\newblock {\em Cold Spring Harbor Perspectives in Biology}, 5(7):a012567, 2013.
\bibitem{Miroshnikov2017}
Alexey Miroshnikov and Matthias Steinr{\"u}cken.
\newblock Computing the joint distribution of the total tree length across loci
  in populations with variable size.
\newblock {\em Theoretical Population Biology}, 118:1--19, 2017.
\bibitem{Myers2005}
Simon Myers, Leonardo Bottolo, Colin Freeman, Gil McVean, and Peter Donnelly.
\newblock A fine-scale map of recombination rates and hotspots across the human
  genome.
\newblock {\em Science}, 310(5746):321--324, 2005.
\bibitem{Kong2010}
Augustine Kong, Gudmar Thorleifsson, Daniel~F. Gudbjartsson, Gisli Masson,
  Asgeir Sigurdsson, Aslaug Jonasdottir, G.~Bragi Walters, Adalbjorg
  Jonasdottir, Arnaldur Gylfason, Kari~Th. Kristinsson, et~al.
\newblock Fine-scale recombination rate differences between sexes, populations
  and individuals.
\newblock {\em Nature}, 467(7319):1099--1103, 2010.
\bibitem{Kong2012}
Augustine Kong, Michael~L. Frigge, Gisli Masson, Soren Besenbacher, Patrick
  Sulem, Gisli Magnusson, Sigurjon~A. Gudjonsson, Asgeir Sigurdsson, Aslaug
  Jonasdottir, Adalbjorg Jonasdottir, et~al.
\newblock Rate of \emph{de novo} mutations and the importance of father's age
  to disease risk.
\newblock {\em Nature}, 488:471--475, 2012.
\bibitem{Jonsson2017}
H{\'a}kon J{\'o}nsson, Patrick Sulem, Birte Kehr, Snaedis Kristmundsdottir,
  Florian Zink, Eirikur Hjartarson, Marteinn~T. Hardarson, Kristjan~E.
  Hjorleifsson, Hannes~P. Eggertsson, Sigurjon~Axel Gudjonsson, et~al.
\newblock Parental influence on human germline \emph{de novo} mutations in
  1,548 trios from {I}celand.
\newblock {\em Nature}, 549:519--522, 2017.
\bibitem{Smith2018}
Thomas C.~A. Smith, Peter~F. Arndt, and Adam Eyre-Walker.
\newblock Large scale variation in the rate of germ-line \emph{de novo}
  mutation, base composition, divergence and diversity in humans.
\newblock {\em PLOS Genetics}, 14(3):1--30, 2018.
\bibitem{Kim2015}
Junhyong Kim, Elchanan Mossel, Mikl{\'o}s~Z. R{\'a}cz, and Nathan Ross.
\newblock Can one hear the shape of a population history?
\newblock {\em Theoretical Population Biology}, 100:26--38, 2015.
\bibitem{Johndrow2018}
James~E. {Johndrow} and Julia~A. {Palacios}.
\newblock Exact limits of inference in coalescent models.
\newblock {\em ArXiv e-prints}, 2017.
\end{thebibliography}

\end{document}